# A universal generic description of the dynamics of the current COVID-19 pandemic

Heinrich Stolz[*,1], Dirk Semkat[2], and Peter Grünwald[3]

[1]Institut für Physik, Universität Rostock, Albert-Einstein-Str. 24, 18051 Rostock, Germany, [2]Institut für Physik, Ernst-Moritz-Arndt-Universität Greifswald, Felix-Hausdorff-Str. 6, 17489 Greifswald, Germany, [3]Institut for Fysik og Astronomi, Aarhus Universitet, Ny Munkegade 120, 8000 Aarhus C, Denmark

*Corresponding author: heinrich.stolz@uni-rostock.de

**Abstract**
The ongoing COVID-19 pandemic is challenging every part of society. From a scientific point of view the first major task is to predict the dynamics of the pandemic, allowing governments to allocate proper resources and measures to fight it, as well as gauging the success of these measures by comparison with the predictions in hindsight. The vast majority of pandemic models are based on extensive models with large numbers of fit parameters, leading to individual descriptions for every hot spot on the world. This makes predictions and comparisons cumbersome, if not impossible.
We here propose a different approach, by moving away from a description over time, and instead choosing the total number of infected people in an enclosed area as the independent variable. Analyzing a few hot spots' data, we derive an empirical formula for the dynamics, dependent only on three variables. The final number of infections is strictly connected to one fit parameter we call mitigation factor, which in turn is mostly dependent only on the enclosed population. Despite its simpleness, this description applies to every of the around 50 countries we have analyzed, allows to separate different waves of the pandemic, provides a figure of merit for the overall usefulness of government measures, and shows when a pandemic is ending. Our model is robust against undetected cases, and allows all nations, in particular those with fewer resources, to reasonably predict the outcome of the pandemic in their country.

**Significance statement**
Based on the empirical data for the number of infections in around 50 countries we propose a simple, yet universally applicable model for the spreading of the COVID-19 pandemic. In contrast to previous models, our description depends on the total number of infected people instead of time. This change of independent variable constitutes the crucial step to analyze different countries worldwide within one mathematical framework with very few parameters. Our model describes the pandemic astonishingly well, including multiple waves, enforcing and loosening of measures and its endpoint, provides a figure of merit to estimate the success of the measures to fight it, and allows all nations, in particular those with fewer resources, to analyze its struggle against the pandemic efficiently.

## I.    Introduction

A proper mathematical modeling is a quintessential prerequisite to understand a pandemic, as was already stated by Sir Ronald Ross [1]. Quite generally, there are two strategies to model pandemics. The "micro"- method relies on following the fate of each member of the population, tracking all its contacts to microscopically model the pandemic in the way of a "Maxwell's Demon" [2]. Unfortunately, just as in the case of the latter concept's origin in thermodynamics, this is not possible due to the lack of sufficient information. Therefore, in the other strategy, only average "macroscopic" aspects, like the total number of infections, or average rates of infection and recovery, are taken into account. Most models are based on the so-called SIR model or extensions from it [1,3-10] and all quantities being described as a function of time. These models divide the population in various compartments as susceptibles (S), infected (I) and recovered (R) in case of the standard SIR model. These compartments are coupled via rates for infection, recovery and mitigation measures (see e.g. [11]) in a set of nonlinear differential equations and their solutions are fitted to the empirical data. An important quantity in these models is the basic reproduction number $R_0$ [4,9,11], since it allows to characterize the dynamics of a pandemic with a single number. Looking at the best available models [11-14] it turns out that the current COVID-19 pandemic requires rather complex extensions of the SIR model with time-varying parameters, the number of which can easily reach more than one hundred. We have tried to reduce the number of parameters as much as possible, but still being able to describe for all countries at least the initial stage of the pandemic, whereby we found a minimum of 18 parameters necessary (see the modelling based on [14] presented in the supplementary materials, section S1).

As a consequence every country which has become a hotspot for the pandemic has been treated individually with hugely varying conclusions (see e.g. [11-28]). This in turn has led to very different approaches to tackle the crisis even in geographically and culturally close countries. Furthermore, the rather abstract nature of the parameters of these models has provoked a lack of public insight undermining the authority of governments worldwide in this time of crisis. A serious problem of this approach, which has been overlooked in the current discussion and might explain the large sensitivity on the free parameters in these models, is that except for the simple SIR model [29], they suffer from chaotic behavior [30].

Going back to mathematics, a pandemic with exponential growth can be described by the first order differential equation [4,5,8,27]

$$\frac{dN(t)}{dt} = \Gamma(t) \cdot N(t) \ , \qquad (1.1)$$

where $\Gamma$ denotes the growth rate which is related to the doubling time $D$ by $\Gamma = \ln(2)/D$, provided that the number of infected persons $N$ is much smaller than total population $N_0 = \text{const}$ (which for the current pandemic is the case). Every equation of this form can be solved exactly if one knows the dependence of the growth rate on time $\Gamma(t)$. Within the SIR type models one tries to obtain this function from solving the pandemic equations. However, another possibility is to know the growth rate $\Gamma(N)$ as function of the number of infections N, given the initial condition N(0) [1]. A well-known example for this is the Gompertz model [5, 8, 31] for which Eq. (1.1) can be cast into a form $dN/dt = r\ln(N_{finit}/N)N$, giving directly $\Gamma(N) = r\ln(N_{finit}/N)$. This model has been used, e.g., to analyze the COVID-19 pandemic in Germany [31] and Iran [32].

The idea central to this contribution is to derive a functional dependence for $\Gamma(N)$ (or equivalently for $D(N)$), which is much more general that the Gompertz model and valid for any pandemic (under the condition of closed populations), from a thorough analysis of the empirical time series of the infection numbers of many realizations of the current pandemic. Obviously, one can characterize with such a general functional relation the dynamics of a pandemic independent from any modelling assumptions. Furthermore, the graphical representation of this relation, which we will call a "characteristic plot", gives intuitive insight in the pandemic.

Such an aim currently seems to be reasonable because of the extreme travel restrictions imposed worldwide, whereby most countries have effectively restricted or completely removed travel across its borders, at least during most of the time of the first wave. This isolation yields the unique advantage of having almost perfectly closed systems with fixed population numbers and supplies a huge number of experimental realizations of the pandemic. In the following, we will demonstrate that by analyzing the empirical data from countries with sufficient resources to give reliable infection numbers (for the problem of undetected infections see section IIIc), to be able to extract a simple law for the general dynamics of the pandemic. This law describes $D(N)$ with only three parameters: 1. The doubling time $D_0$ at the beginning of the pandemic characterizing the initial spread of the infection, 2. a proportionality constant $\rho$ coined "mitigation factor" that describes an exponential growth of $D$, and 3. the end point of the pandemic $N_{finit}$. Most important, we find a strict correlation between $\rho$ and $N_{finit}$, the smaller $\rho$ the larger $N_{finit}$. Based on this very simple model we show that all centers of outbreak follow this law. Furthermore, it allows us to identify different waves in specific countries as well as the onset and easing of measures, evaluate the usefulness of certain measures, and, most importantly, provide a criterion for when the epidemic ends. All of this allows drawing very general conclusions on how the pandemic evolves in different countries, how to act now, when a new wave of outbreaks hits or a new pandemic arises.

---

[1] The implicit solution: $t = \int_{N(0)}^{N} dx/(x\Gamma(x))$ can be obtained by the method of separation of variables [5].

## II. Results

For a first test we choose four countries with a wide spread in population numbers namely Luxembourg (LU), Switzerland (CH), the Netherlands (NL), and Italy (I). In all countries we can expect the data to be reliable. The infection data were taken from [33,34]. Given a time series of infection data $\{t_i, N_i;\ i=1\ldots M\}$, an estimate for $\Gamma(t)$ can be calculated by $\Gamma(t_i) \approx (N_{i+1} - N_i)/[N_i \cdot (t_{i+1} - t_i)]$. From these the dependence of $\log(D(N))$ on $N$, which will be called the "characteristic plot" can be calculated and is shown in Fig. 1. We can clearly identify four phases: A) an initial phase with very few infection numbers (marked by the black dots) with irregular behavior, probably due to statistical error because of the small numbers, which we will neglect. B) A second phase ("linear regime"), where the doubling times increase exponentially (linear increase in the logarithmic plot) marked by red dots. C) The exponential increase is followed by a super-exponential rise (blue dots), which however stops before diverging (which would be the endpoint of the pandemic $N_{finit}$). D) The last phase (marked by magenta points) shows a decrease of $D(N)$. Of course it would be possible to describe this phase by a negative slope of $\log(D(N))$. However, as suggested by the time dependence of the daily numbers of new infections that clearly show a second wave, a straightforward consideration shows that this phase characterizes indeed a "second wave" of the pandemic. Because of the statistical independence of the different waves, one can assume that for $N > N_{finit}$ only the second wave is active. Therefore, its characteristic plot is obtained by subtracting from the total infection numbers just $N_{finit}$ (or to be more accurate -as has been done in the actual analysis- the infection numbers N(t) calculated from Eq. (1.1)) and calculating from these data the doubling times, which are given in Fig. 1 by the orange filled circles, which show a similar behavior as in the first wave. To describe then another wave of the pandemic one has to set up another differential equation (1.1) and integrate it with the then applicable mitigation factor and endpoint. The total infection number is then the sum of both partial infection numbers.

From this analysis, we suggest the following "ansatz" for the growth constant $\Gamma$ (during one wave of a pandemic)

$$\Gamma(N) = \frac{\ln(2)}{D_0} \exp(-\rho_0 \cdot N) / f_{\text{finit}}(N) \ . \tag{1.2}$$

The constant $\rho_0$ will be designated as the "mitigation factor" as it reflects the effectiveness of mitigation measures and $D_0$ the initial doubling time. The function $f_{\text{finit}}(N)$ describes a super-exponential growth of $D$ and can be assumed as[2]

$$f_{\text{finit}}(N) = 1 + \frac{1}{25} \tan^2\left(\pi \frac{N}{2 N_{\text{finit}}}\right) \ . \tag{1.3}$$

To obtain the parameters $D_0$, $\rho_0$ and $N_{\text{finit}}$ we applied a least-square fit to the empirical doubling times in the first and second wave of the pandemic. Note that for the Netherlands, Switzerland, and Italy, the second wave is still in its "linear" regime and here we are not able to deduce the endpoint. The results are given in Table 1 and from these we generate the blue lines in Fig. 1. For a straightforward solution of Eq. (1.1) (together with Eq. (1.2)) with a differential equation solver we need to specify only the number of infections $N_{\text{start}}$ at a certain date $t_{\text{start}}$ where we set the origin of time to the 1st February 2020, and used the parameters $D_0$, $\rho_0$ and $N_{\text{finit}}$ from table 1. The results are depicted in Fig. 2 showing that the empirical data are extremely well described by our model (red lines) for all phases of the pandemic. Notice that the noise and also the oscillations present in the empirical data have only little influence on the result of the integration as they average out.

Exemplarily, we look more closely on the time dependence of the doubling time for Switzerland (green and magenta circles and lines). In the first days of the pandemic the doubling times are almost constant (with some statistical jitter due to the small numbers) leading to the exponential growth of the infection number. Then the measures used by all countries to control the epidemic begin to work and the doubling times increase up to a maximum and then decrease again. We interpret this decrease not as a change in mitigation measures but as the beginning of a second wave following the same rules as the first wave.

---

[2] This function has been chosen because it describes the empirical data rather well, but using any other function with a divergence at $N = N_{\text{finit}}$ is possible.

Note the quite substantial non-linearity, despite the simplicity of the laws given by Eqs. (1.2) and (1.3).

Since the initial doubling time $D_0$ should depend only on genuine properties of the pandemic, we expect it to be almost constant and indeed this is the case. In contrast, $\rho_0$ and $N_{\text{finit}}$ are varying considerably with the number of inhabitants of the country chosen, which were taken from [36]. However, despite of the huge variation of the population numbers by two orders of magnitude we find that the product of the mitigation factor and the end number of infections turns out to be almost constant (deviations less than 10%), see the last entry of Table 1. This means that the end of the pandemic is closely related to the exponential increase of doubling times with infection number suggesting a relation

$$N_{\text{finit}} = c \cdot \rho^d . \qquad (1.4)$$

We also find that the product of population number and mitigation factor $N_0 \cdot \rho_0$ remains nearly constant (see Table 1), however, with a larger variation. This suggests a relation of the form

$$\rho = a \cdot N_{\text{pop}}^b . \qquad (1.5)$$

If both relations can be substantiated from our analysis, this would have far reaching consequences, because it means that the total number of infections depends only on the number of inhabitants in a certain country!

To check whether these relations are indeed a generic behavior of the pandemic, we applied our analysis to a large number of other countries, i.e. Australia (AU,*), Austria (AT), Belgium (BE,*), Brazil (BR,*), Czechia (CZ), Denmark (DK), France (FR), Germany (DE,*), Hungary (HU), Iceland (IS), India (IN,*), Ireland (IE), Japan (JP), New Zealand (NZ), Russia (RU), South Korea (KR,*), Spain (ES,*), Sweden (SE,*), Taiwan (TW,*), United Kingdom (UK,*), United States of America (US), and also to several states of Germany like Bavaria (Bay), Hamburg (HH,*) North Rhine-Westphalia (NRW) and Mecklenburg-Western Pomerania (MV,*). For a selected number of cases (marked by an *) the detailed fits of the doubling times and the solutions of Eq. (1.1) are plotted in Fig. S2 to S8 (see SOM). For all countries the parameters of the fits and the number of inhabitants in each country or province can be found in Table S1. One should note that most of the countries follow the regular behavior shown in Fig. 1: an exponential increase of the doubling time followed by a super-exponential growth, eventually followed by a second wave. However, there are some special cases, which on first sight do not follow our model. However, as shown in Section IIIb, this is not the case. All these countries will be indicated with blue color in Fig. 3 where in part **a** the endpoints are plotted against the mitigation factor, in part **b** the latter is plotted against the number of inhabitants. The initial doubling times are plotted in Fig. S1. In all cases, if not otherwise marked, the parameters from the first wave are taken.

### III. Discussion
#### a) Derivation of generic laws

In Fig. 3a the countries with regular behavior given by the full triangles obviously following very good our ansatz Eq. (1.4), the regression giving $c = 2.987 \pm 0.017$ and $d = -1.015 \pm 0.013$. The open blue triangles denote those countries which are still in the exponential growth regime, whereby the endpoints calculated with Eq. (1.4) have been used as ordinates. Note that also countries, which are commonly considered as "special" cases like NZ, TW, SK, or SE satisfy the generic law Eq. (1.4).

In Fig. 3b the mitigation factors show much more variation between the different countries as suggested by Eq. (1.5), which means that the main factor that determines the pandemic is the number of inhabitants in each country. That there have to be other influencing effects is obvious, as otherwise all points in Fig. 3b would lie on the same line. One of such factors might be the population density of the region considered. That this is indeed influencing the mitigation factor can be seen by a comparison of MV with HH (see table S1) which have almost the same population with a ratio of 35 in population density, but the mitigation factors differ by a factor of 6.5. Therefore this must be of minor importance (see also Fig. S9).

Therefore, we interpret the deviations as stemming from all the different measures that have been used to mitigate the pandemic, from complete lock-down (like in China) to recommendations to reduce social life and take care like in Sweden. To identify such effects a detailed multivariate study of all the measures has to be done, which however, is far outside the present contribution and left as interesting work for the future.

Here we only want to remark that the countries fall into two groups (denoted by full and open red triangles), which behave quite similar with respect to the population numbers. The full and dashed lines are the result of a logarithmic regression of Eq. (1.5) giving the following parameters: $a = 1282 \pm 75$ (with $a = 117 \pm 30$ for the dashed line) and $b = -1.12 \pm 0.06$.

Especially interesting for future studies are countries with more than one wave that switch between these groups (as AU or IN).

The combination of Eq. (1.5) and (1.4) indeed shows that the maximum final number of infections and the number of inhabitants of a country is closely correlated, provided that the country uses the standard measures of border control and social distancing. The expected endpoint of the pandemic is given by $N_{\text{finit}}^{\text{pred}} = c \cdot a^d \cdot N_{\text{pop}}^{b \cdot d}$ (in the case where the endpoint is not reached, one may take as a guess the endpoint calculated with Eq. (1.4)). We can therefore define as figure of merit for the effectiveness of pandemic control the ratio

$$M = \frac{N_{\text{finit}}^{\text{pred}}}{N_{\text{finit}}} = \frac{c\, a^d\, N_{\text{pop}}^{b \cdot d}}{N_{\text{finit}}}, \qquad (1.6)$$

which is plotted in Fig. 4. Obviously, countries with $M \gg 1$ have done an excellent job in controlling the pandemic, those with $M < 1$ have failed in this respect. Comparing with Fig. 3b we can state that countries with high mitigation factor (near the dashed line) have a very large $M$, while countries with mitigation factor below the full line in Fig. 3b show a worse performance in the pandemic.

### b) Exceptions and multiple waves

While the pandemic in the majority of the investigated countries follows closely our model, there are a few cases, where our model seems to be not applicable. However, every exception we found could be explained as one of the following two cases:

1. The pandemic shows phases with different (positive) mitigation factors. Examples are BR, DK, IN, RU, US. As shown for the case of Denmark in Fig. 5a, (see also the fits in Figs. S2 and S3) the pandemic can still be described by our ansatz, but with a multi-mode behavior where

$$\Gamma(\nu N_0) = \sum_i \Gamma_{0i} \exp(-\rho_{0i} \cdot \nu N_0)\Theta(\nu_{bi} - \nu)\Theta(\nu - \nu_{ei}) / f_i(\nu N_0), \qquad (1.7)$$

with $\Theta(x)$ being the Heaviside step function and $\nu_{bi}, \nu_{ei}$ denoting begin and end of the $i^{\text{th}}$ period of the pandemic, respectively. In other words, our model clearly detects different phases of the pandemic, which possibly are related to onset and loosening of measures in the respective regions.

2. The pandemic shows a phase where the doubling times decrease with infection numbers, but does not show multiple peaks in the daily infection numbers. Examples are Sweden, South Korea, and the USA. Here, as seen from the analysis in case of USA (Fig. 5b), a second (or even third) wave of the pandemic is hidden in the dynamics (see SK in Fig.S3). It is here even possible to determine the endpoint of the first wave. Actually, every decrease of the doubling time has to be explained by the onset of a second independent wave.

What is most important for the community is that our model can be applied easily to the world as a whole. This is shown in Fig. 6, where the characteristic plot (see inset) shows that the pandemic exhibits clearly a two phase behavior, probably related to a phase were the pandemic has been effective mostly in Europe and later a phase where it was active in America and India. At the moment, the pandemic is still in the "linear" regime, with the endpoint too far to be detectable. However, from the mitigation factor and relation (1.4) one can estimate that in the end about 90-100 Millions of people will be infected. Note that this figure also demonstrates the predictive power of our model, as the modelling has been done only from the empirical data up to $1^{st}$ of July, and the agreement with the data up to August, $15^{th}$ is almost quantitative.

### c) undetected infections

Finally, we shortly discuss the problem of correct infection numbers which is inherent in all modelling studies. Increasing the number of infections by $\varepsilon$, the daily doubling time is not altered, only the mitigation factor is reduced by $1/(1+\varepsilon)$. By this only the position of the country in Fig. 3b is changed. However, the important relation between the endpoint and mitigation factor (Fig. 3a) is not changed at all. This means that our model is quite robust against uncertainties in the exact number of infections.

In conclusion, we have shown that the current COVID-19 pandemic follows generic laws that can be described by three parameters, the initial doubling time $D_0$, the mitigation factor $\rho$, and the endpoint of the pandemic $N_{finit}$, whereby the latter two are found to be inversely proportional. While the initial doubling time is almost the same for all countries investigated, the mitigation factor and the endpoint, however, depend mainly on the number of inhabitants of the area that is closed during the pandemic, and which is mostly identical with the population of a certain country. Therefore, our model allows the qualitative prediction of the pandemic for any country only from the number of inhabitants. Furthermore, by calculating the figure of merit $M$ (Eq. (1.6)) one can assess the success of the measures taken to reduce the pandemic. Finally, we stress that we only considered the number of infections. Whether the death toll also can be described by a similar generic law requires an independent study.

**Methods**

We used for our analysis of the pandemic the numbers provided by the Johns Hopkins University [33] up to the 14[th] of August 2020. We checked the reliability of the data by comparing to the list provided by the "worldometer" organization [34]. The data for the German states were obtained from the daily lists of the German newspaper "Die Morgenpost" [35]. The latter data can be found in the Supplementary Materials. All others are available at the corresponding web pages. The mitigation factor $\rho$ and the endpoint of the pandemic were obtained by a least square fitting of the logarithm of the doubling times to Eq. (1.2) using standard statistical methods. The same has been done with the regression of Eqs. (1.5) and (1.4) to obtain the parameters $a, b, c,$ and $d$. As differential equation solver we used the routines from MATHCAD15 [37], an example can be found in the supplementary material.

**Acknowledgements:** The authors thank W.-D. Kraeft, University of Greifswald, and S. O. Krüger, University of Rostock, for helpful discussions. We thank R. Walke, MPI for Demographic Research, Rostock for helpful comments and especially for coining the word "mitigation factor". D.S. thanks the Deutsche Forschungsgemeinschaft for financial support (project number SE 2885/1-1).

**Funding:** The authors H.S. and P.G. did not receive any funding. **Author contributions:** The authors contribute equally to the paper.

**Competing interests:** The authors declare no competing interests.

**Materials and correspondence:** Available from Corresponding author: heinrich.stolz@uni-rostock.de

**Supplementary Material**

Section S1

Figures S1-S4

Tables S1

Data file Data_corona1.xls

Program file Switzerland_cor.pdf

| Country | Inhabitants in millions $N_0$ | $D_0[d]$ | $10^4 \cdot \rho_0$ | $N_{\text{finit}}$ | $10^{-3} \cdot N_0 \cdot \rho_0$ | $N_{\text{finit}} \cdot \rho_0$ |
|---|---|---|---|---|---|---|
| Luxembourg (LU) | 0.63 | $1.35 \pm 0.22$ $4.66 \pm 0.26$ | $0.974 \pm 0.79$ $0.637 \pm 0.38$ | $4150 \pm 75$ $5000 \pm 500$ | $0.624 \pm 0.036$ $0.400 \pm 0.024$ | $4.04 \pm 0.15$ $3.17 \pm 0.32$ |
| Switzerland (CH) | 8.65 | $2.01 \pm 0.22$ $4.56 \pm 0.26$ | $1.008 \pm 0.049$ $2.641 \pm 0.241$ | $31580 \pm 150$ - | $0.872 \pm 0.03$ $2.26 \pm 0.21$ | $3.185 \pm 0.11$ - |
| Netherlands (NL) | 17.13 | $3.14 \pm 0.11$ $4.38 \pm 0.23$ | $0.672 \pm 0.017$ $1.002 \pm 0.125$ | $52860 \pm 200$ - | $1.151 \pm 0.030$ $1.716 \pm 0.214$ | $3.76 \pm 0.10$ - |
| Italy (I) | 60.46 | $3.01 \pm 0.07$ $14.64 \pm 1.38$ | $0.1406 \pm 0.0019$ $0.16 \pm 0.22$ | $(2.485 \pm 0.02) \times 10^5$ - | $0.850 \pm 0.011$ $0.97 \pm 1.24$ | $3.494 \pm 0.047$ - |

**Table1.** Model parameters for the four countries as obtained from the fits in Fig. 1A. The second column gives the number of inhabitants in 2020, the third the initial doubling time, the fourth the mitigation factor, and the fifth the number of infected people at the end of the pandemic. The sixth and seventh columns give the products of mitigation factor with the population number and with the number of infected people at the end of the pandemic.

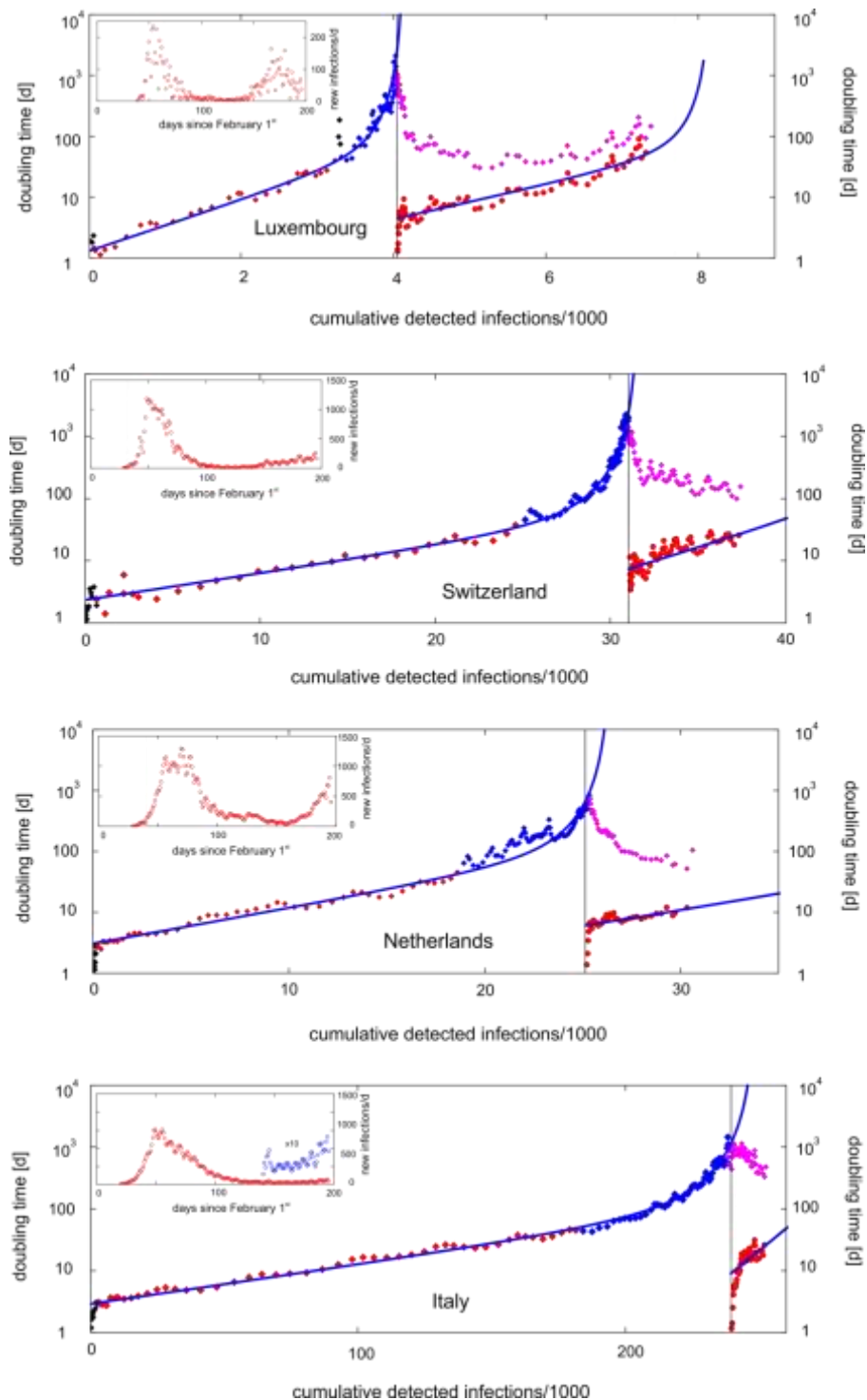

**Fig. 1.** Daily doubling times as calculated from the number of infections vs. cumulative number of infected persons for four countries (top/down: Luxembourg, Switzerland, Netherlands, Italy). The data show clearly a four-phase behavior, first (black dots) an irregular behavior at the beginning due to statistical errors, second an exponential increase (red dots) followed by a super-exponential behavior (blue dots) peaking at some critical infection number (vertical black line), and followed by a phase with decreasing doubling time (magenta dots). The red dots after the black vertical line are doubling times calculated by subtracting cumulative infection number of the first wave from the data (see text). The insets show the daily infection numbers showing in all cases a second wave.

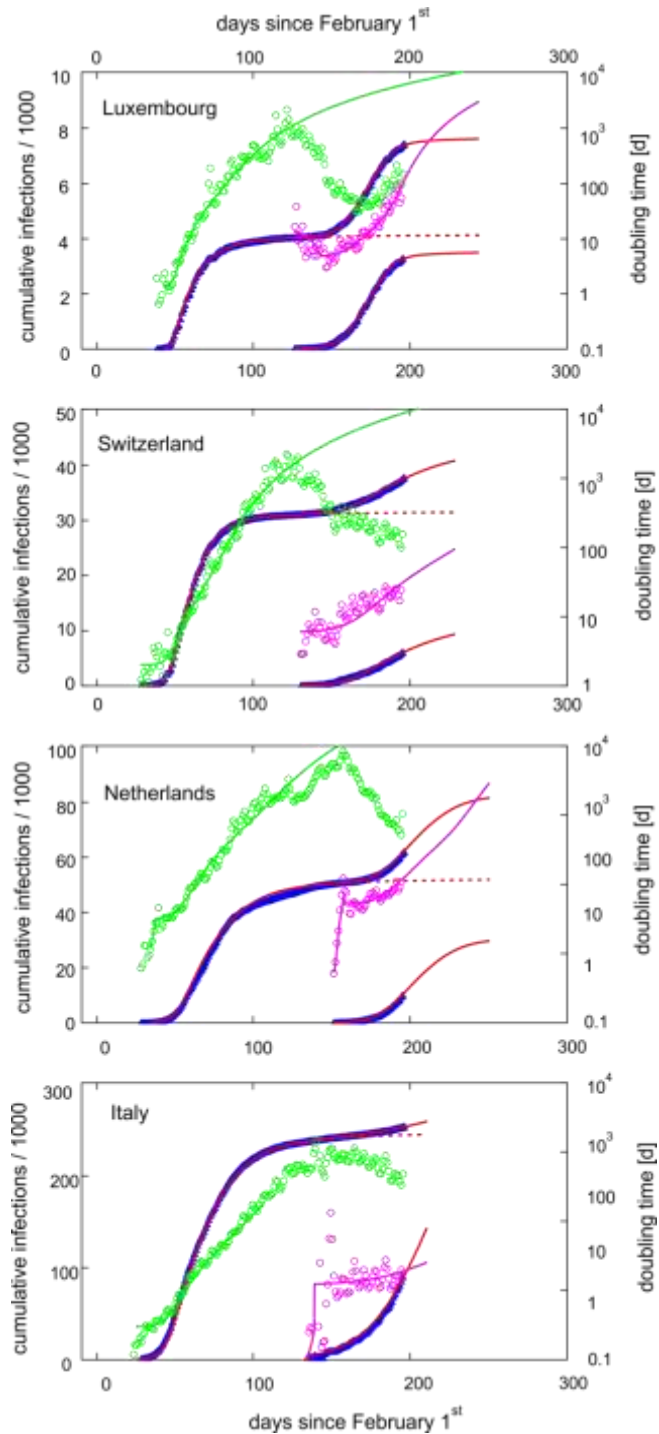

**Fig. 2.** Time dependence of cumulated infection numbers for the four countries (top/down: Luxembourg, Switzerland, Netherlands, Italy). Blue triangles: empirical data (I: total infections, II: only second wave). The red lines show the results obtained from solving Eq. (1.1) and (1.2) using the parameters from Table 1. dotted for the first wave, full total infection numbers (I), only second wave (II). The open green circles (and lines) show the daily doubling times only for the first wave, magenta circles and lines that for the second wave (right ordinate).

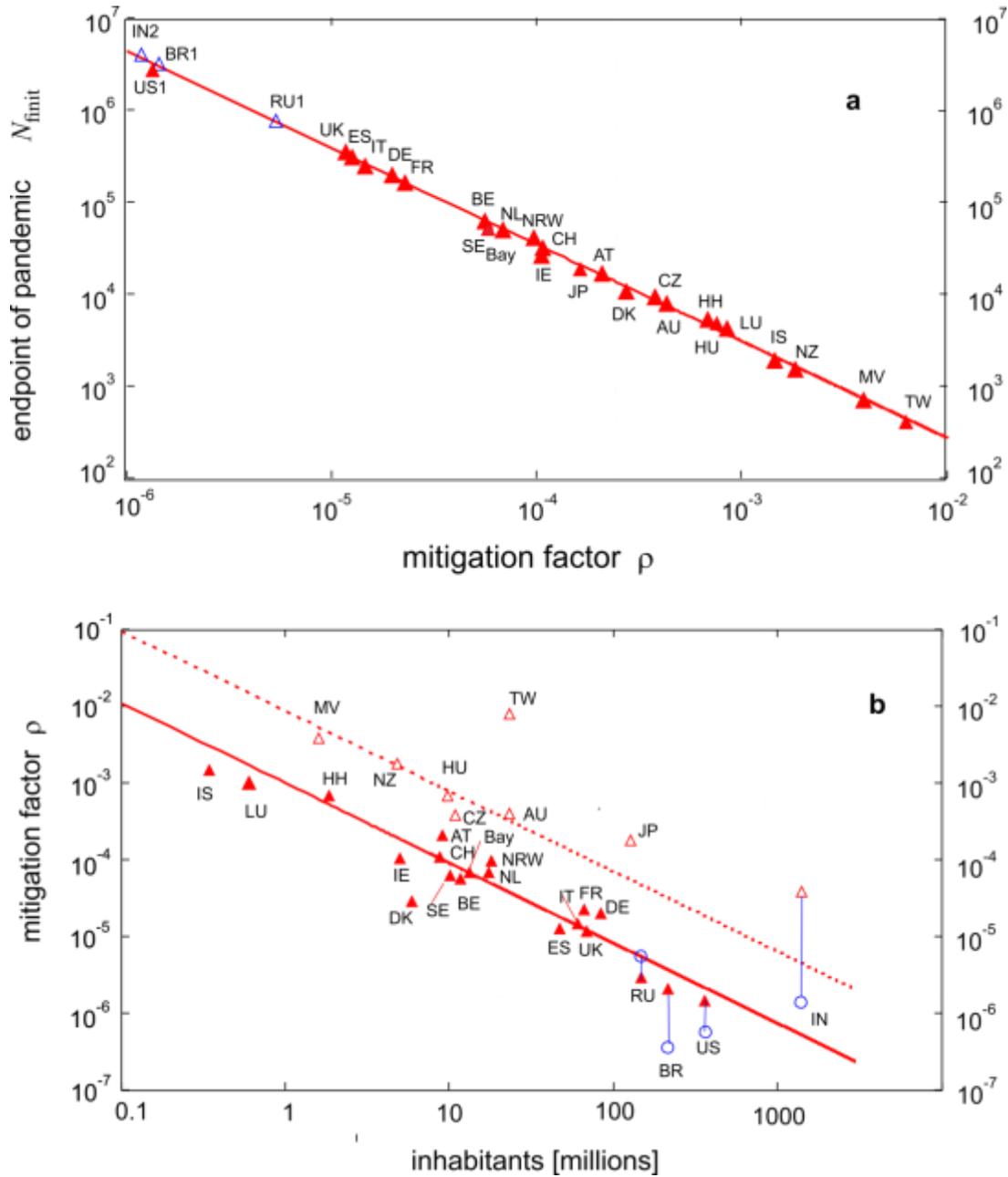

**Fig.3** Part a: Endpoints of the pandemic vs. the mitigation factor. The full red triangles denote countries with a clear transition into a super-exponential increase of the doubling times (phase 3), which have been included into a logarithmic regression with Eq. (1.4). The blue open triangles denote countries which are still in the "linear" regime. Here the expected endpoints are calculated using Eq. (1.4). The full list of countries and abbreviations is given in Table S1.

Part b: Mitigation factors $\rho$ vs. the number of inhabitants for various countries. The red triangles are those countries, which have been included in the logarithmic regression with Eq. (1.5) giving the full line, the open red triangles are the countries, which deviate strongly from the "normal" behavior and have been omitted in the regression. The blue circles denote the countries with a two phase behavior (second phase). These points are connected by blue lines to the results of the corresponding first phase.

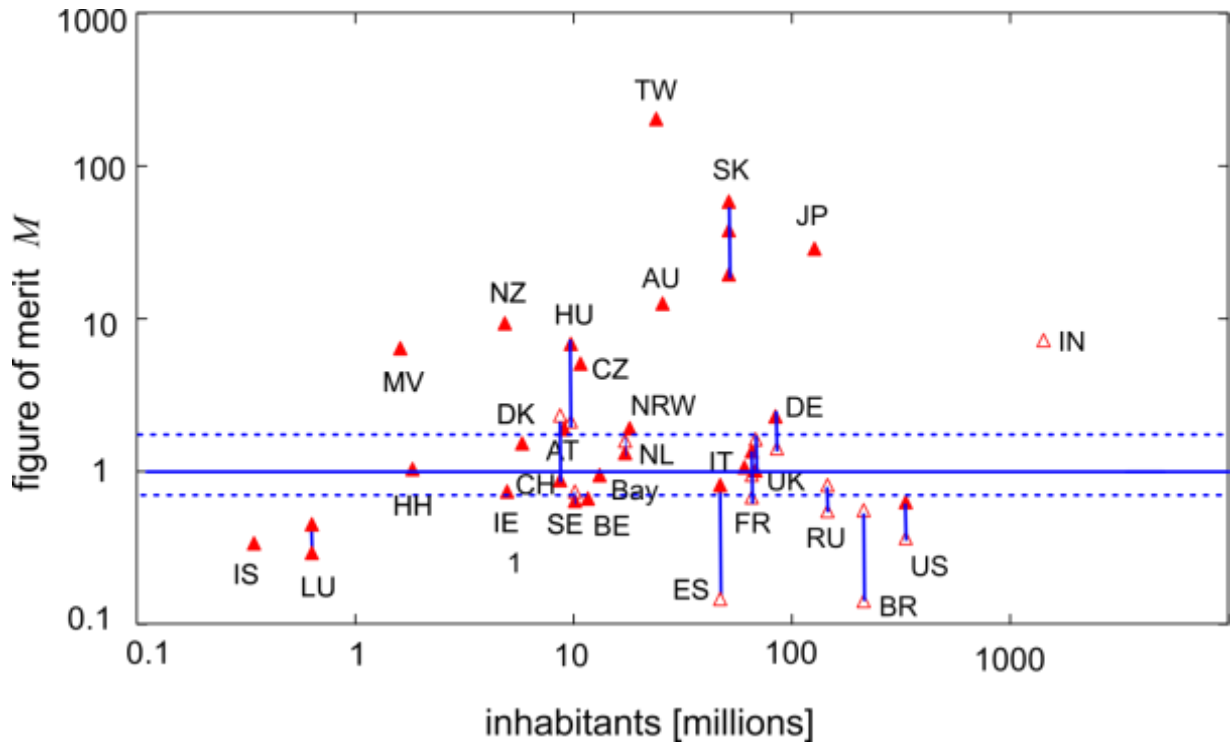

**Fig. 4.** : Figure of merit $M$ (see Eq. (1.6)) for the countries investigated. The full line indicates $M=1$, the dashed lines give the standard deviation of $M$ from the statistical errors in the parameters a,b,c and d. The full symbols denote those countries with a complete first wave of the pandemic, the open symbols denote countries which are still in the "linear phase" of a second or third wave (like FR or US) or in the linear phase of the first wave (like IN or BR). The vertical blue lines connect the different waves (or phases) of the same country.

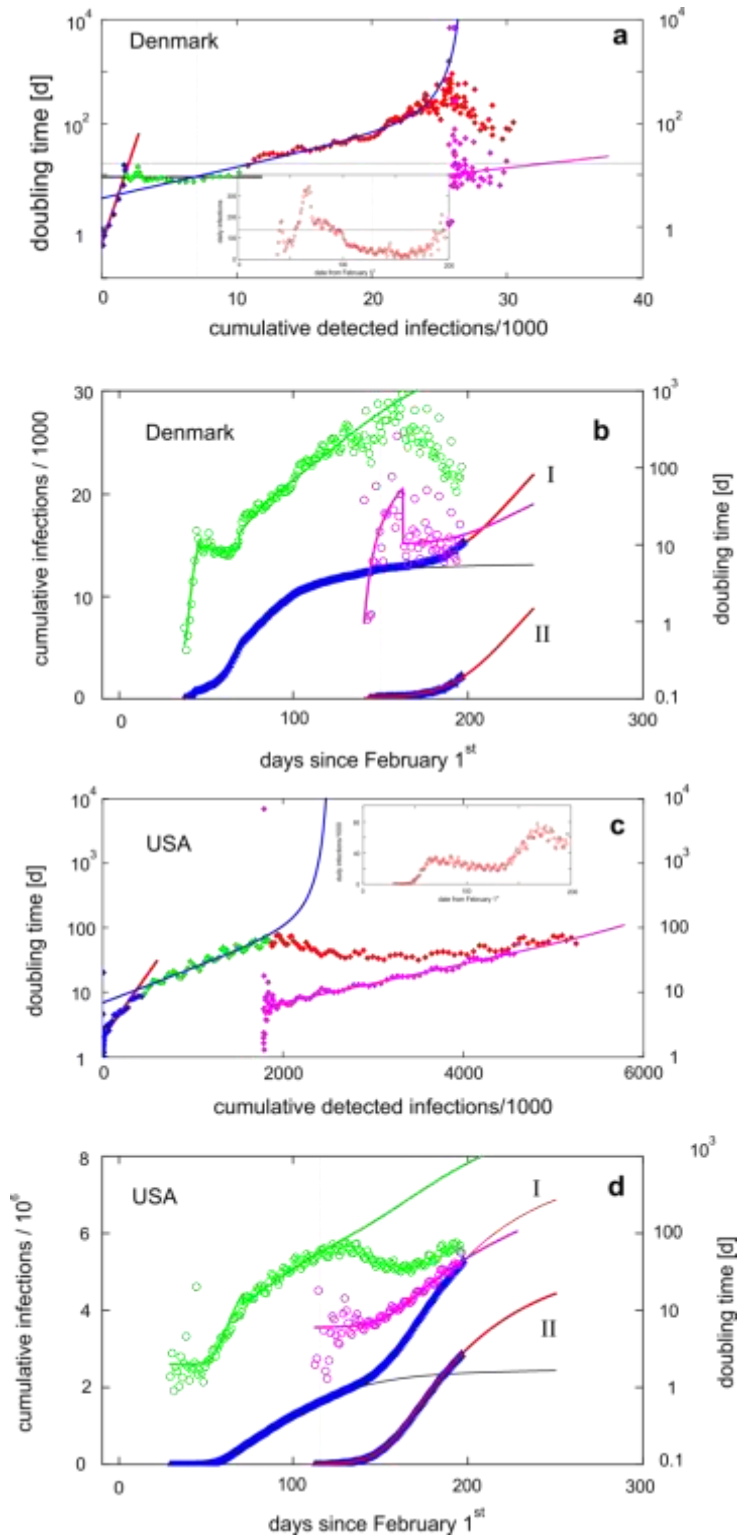

**Fig. 5.** Analysis of pandemic for Denmark (a,b) and the USA (c,d). Plotted are the daily doubling times as calculated from the number of infections vs. cumulative number of infected persons (a,c) in the same way as in Fig.1, the insets show the daily infection numbers. Parts b and d show the time dependence of cumulated infection numbers. The blue triangles show the empirical data (I: total infections, II: only second wave). The full lines show the results obtained from solving Eq. (1.1) and (1.2) using the parameters from Table S1: gray for the first wave, red total infection numbers (I), or only the second wave (II). The open green circles (and lines) show the daily doubling times only for the first wave, magenta circles and lines that for the second wave (right ordinate).

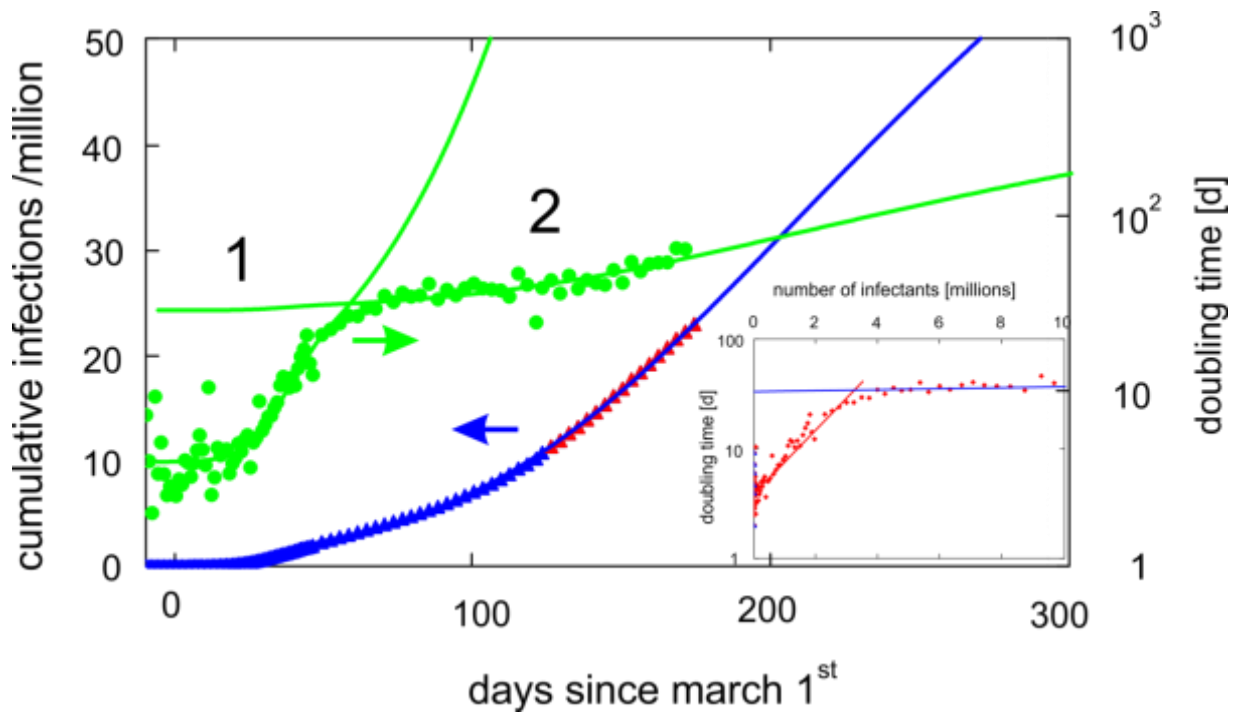

**Fig. 6**: Analysis of the world-wide COVID-19 pandemic with our generic model. The panel shows the time dependence of the infection numbers, blue triangles mark the empirical data, which have been used in the determination of the model parameters $D_{0i}, \rho_{0i}$ for the two phases 1 and 2 by fitting the dependence of the doubling times on the infection numbers (inset) , the full blue line gives the result of our model using the hypothetical $N_{finit2}$ = 95 millions calculated from Eq.(1.4), $N_{finit1}$ =10 millions. The green dots gives the doubling times with the green lines showing the result of the calculation, the phases 1 and 2 are marked.

# Supporting Information

for

"A universal generic description of the dynamics of the current COVID-19 pandemic

by

Heinrich Stolz, Dirk Semkat, and Peter Grünwald

Table S1

Figs. S1-S4

Text S1: Analysis of Pandemic by SIR-based models

Example program file "Mathcad - Corona_analysis_Switzerland.pdf"

Table of data for German states "corona_data_hh_mv1.pdf"

| country | Short notation | Inhabitants $N_0$[millions] | $D_0[d]$ | $\rho_0$ | $N_{\text{finit}}$ | $10^3 N_0 \rho_0$ | $10^3 N_{\text{finit}} \rho_0$ |
|---|---|---|---|---|---|---|---|
| Australia | AU | 25.5 | 1.67 | 4.30e-4 | 7200 | 12.32 | 3.478 |
| Austria | AT | 9.01 | 1.91 | 2.072e-4 | 1.70e4 | 1.867 | 3.523 |
| Bavaria | Bay | 13.1 | 2.613 | 6.81e-5 | 4.92e4 | 0.892 | 3.351 |
| Belgium | BE | 11.59 | 3.04 | 5.49e-5 | 6.3e4 | 0.636 | 3.456 |
| Brazil | BR | 212.6 | 7.36 | 1.772e-6 | - | 0.377 | - |
|  |  |  | 13.9 | 4.517e-7 | - | 0.096 | - |
| Czechia | CZ | 10.71 | 2.40 | 4.59e-4 | 9000 | 0.636 | 4.131 |
| Denmark | DK | 5.79 | 3.5 | 2.74e-4 | 1.35e4 | 1.586 | 3.075 |
| France | FR | 65.27 | 2.93 | 1.62e-5 | 1.90e5 | 1.057 | 3.077 |
|  |  |  | 20 | 8.10e-6 | - | 0.529 | - |
|  |  |  | 5.7 | 1.134e-5 | - | 0.740 | - |
| Germany | DE | 83.78 | 2.32 | 2.073e-5 | 1.98e5 | 1.651 | 4.105 |
|  |  |  | 14 | 1.348e-5 | - | 1.129 | - |
| Hamburg | HH | 1.82 | 2.2 | 6.815e-4 | 5.30e3 | 1.240 | 3.612 |
| Hungary | HU | 9.66 | 3.98 | 6.949e-4 | 4.3e3 | 6.95 | 3.093 |
|  |  |  | 3.0 | 2.158e-4 | - | 2.09 | - |
| Iceland | IS | 0.342 | 2.369 | 1.45e-3 | 1.9e3 | 0.496 | 2.755 |
| Ireland | IE | 4.94 | 1.572 | 1.59e-4 | 2.64e4 | 0783 | 4.197 |
| India | IN | 1380 | 9.256 | 2.842e-6 | - | 3.922 | - |
| Italy | IT | 60.46 | 3.197 | 1.384e-5 | 2.485e5 | 0.837 | 3.3439 |
| Japan | JP | 126.5 | 3.414 | 1.635e-4 | 1.71e4 | 20.68 | 2.796 |
| Luxembourg | LU | 0.63 | 1.35 | 9.74e-4 | 4150 | 0.635 | 4.041 |
|  |  |  | 4.74 | 6.33e-4 | 5500 | 0.399 | 3.50 |
| Mecklenburg-W. Pomerania | MV | 1.6 | 2.0 | 4.86e-3 | 820 | 0.843 | 3.984 |
| Netherlands | NL | 17.13 | 3.14 | 7.054e-5 | 5.33e4 | 1.208 | 3.76 |
|  |  |  | 4.71 | 8.464e-5 | - | 1.45 | - |
| New Zealand | NZ | 4.82 | 1.514 | 2.063e-3 | 1.55e3 | 9.943 | 3.198 |
| North-Rhine-Westphalia | NRW | 18 | 2.561 | 9.669e-5 | 4.15e4 | 1.74 | 4.303 |
| Russia | RU | 145.2 | 5.84 | 3.978e-6 | - | 0.581 | - |
|  |  |  | 11.3 | 2.682e-6 | - | 0.392 | - |
| South Korea | SK | 51.27 | 1.216 | 3.047e-4 | 8400 | 15.622 | 2.559 |
|  |  |  | 2.5 | 9.141e-4 | 2500 | 46.865 | 2.285 |
|  |  |  | 8 | 5.942e-4 | 5400 | 30.462 | 3.208 |
| Spain | ES | 46.75 | 2.214 | 1.412e-5 | 2.45e5 | 0.660 | 3.46 |
|  |  |  | 10.0 | 2.543e-6 | - | 1.188 | - |
| Sweden | SE | 10.1 | 6.522 | 6.19e-5 | 5.8e4 | 0.625 | 3.59 |
|  |  |  | 3.5 | 7.12e-5 | - | 0.719 | - |
| Switzerland | CH | 8.65 | 2.165 | 1.008e-4 | 3.16e4 | 0.872 | 3.185 |
|  |  |  | 6.0 | 2.671e-4 | - | 1.192 | - |
| Taiwan | TW | 23.8 | 1.718 | 7.503e-3 | 455 | 0.179 | 3.414 |
| United Kingdom | UK | 67.89 | 3.828 | 1.161e-5 | 2.84e5 | 0.788 | 3.295 |
|  |  |  | 10.0 | 1.858e-5 | - | 1.261 | - |
| USA | US | 331 | 6.898 | 1.219e-6 | 2.989e6 | 0.403 | 3.643 |
|  |  |  | 5.5 | 7.109e-7 | - | 0.235 | - |

**Table S1.** Parameters for all countries investigated. The second column gives the shortcuts used in Fig.3, 4, and S1, the asterix mark the countries displayed in Fig. S2 to S4. The third column gives the number of inhabitants taken from [36].The fourth columns gives the initial doubling time, the fifth the mitigation factor and the sixth the endpoint of a pandemic wave (a dash denotes the cases where the endpoint is not in sight). The seventh column gives the product of the mitigation factor with the number of inhabitants, while the eights denotes the product of the mitigation factor with the with the endpoint of a pandemic wave.

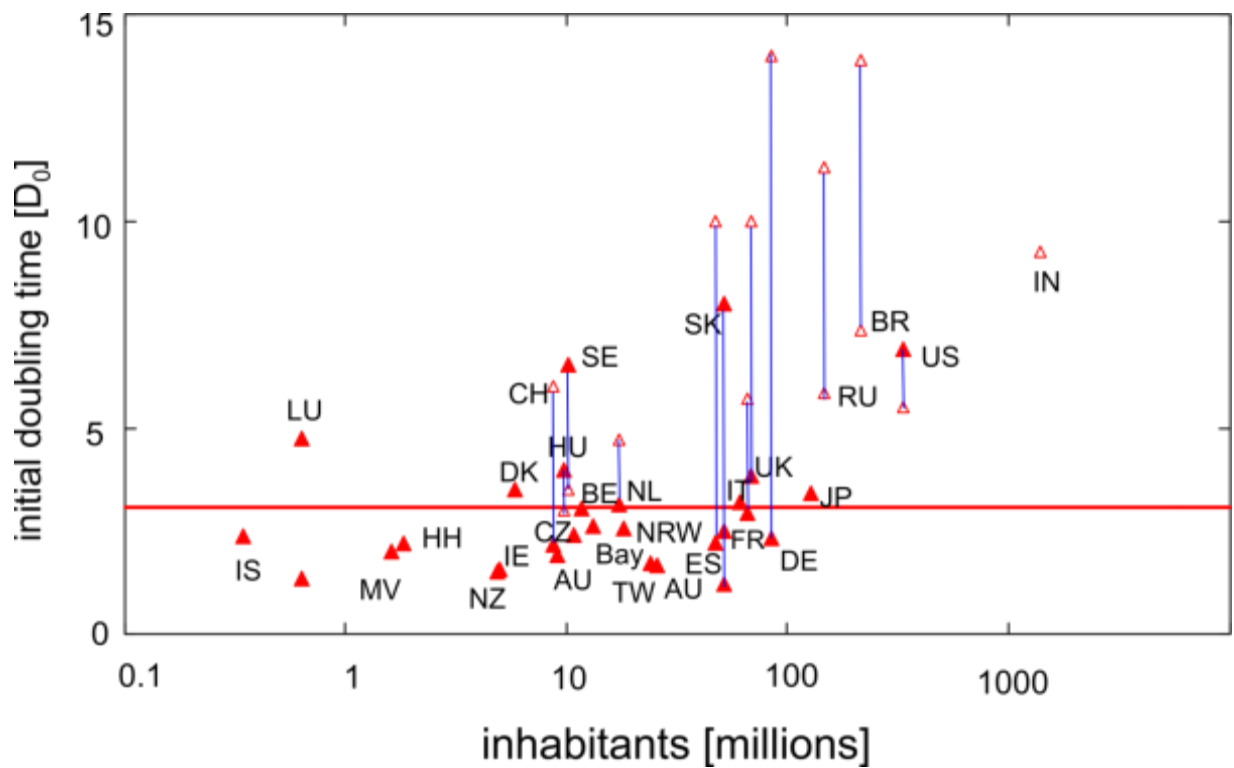

**Fig. S1.**
The initial doubling times vs. population numbers for the countries shown in Fig. 2. The full symbols denote countries with a completed first wave, the open symbols denote either countries with a second wave (connected by a blue vertical line to the first wave data) or those still in the linear region. The red line gives the average value of $D_0 = 3.02 \, \text{d}$.

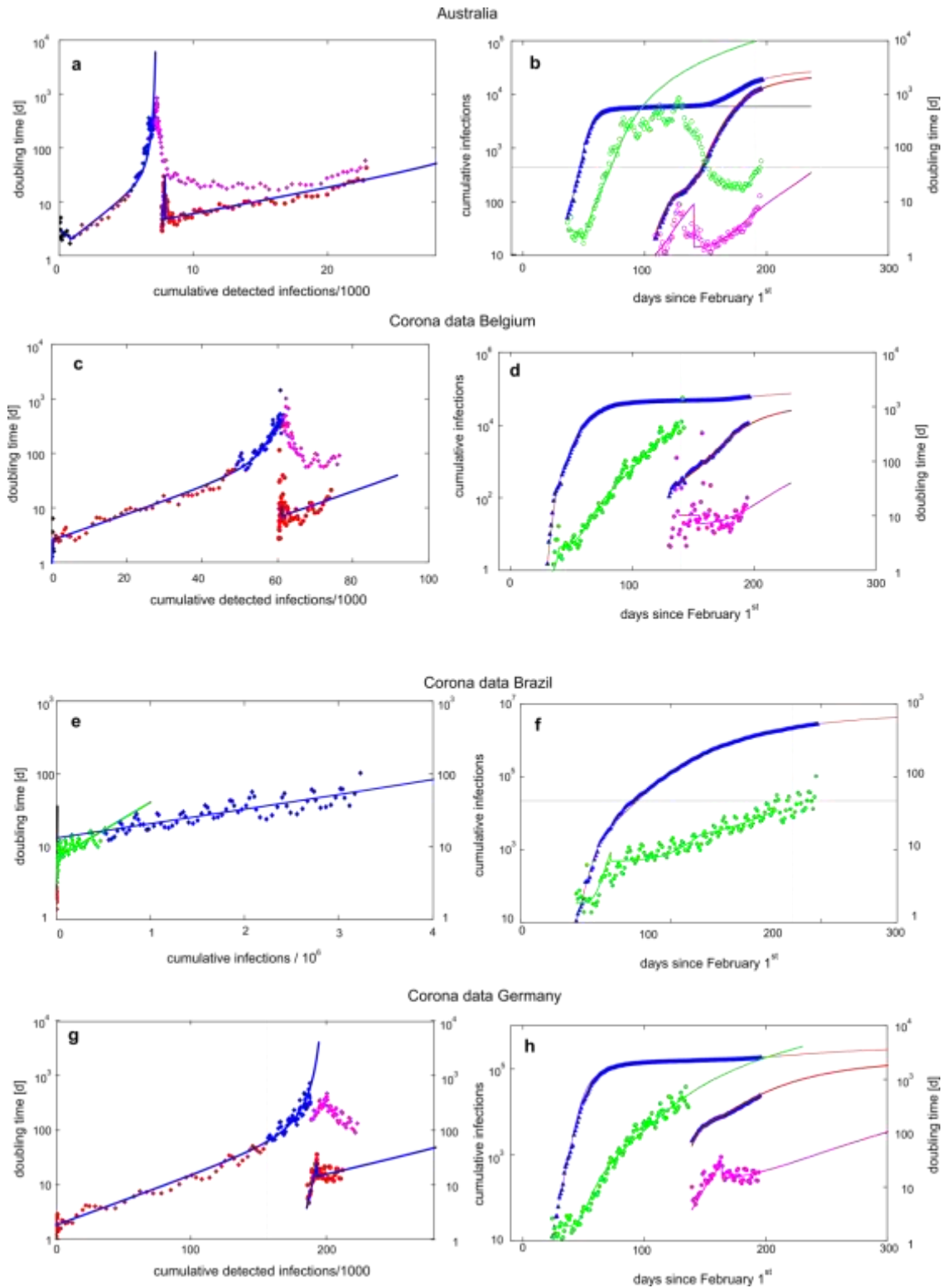

**Fig. S2.** Left side: Daily doubling times as calculated from the number of infections vs. number of infected persons for the indicated countries. Blue triangles: data used to fir the growth factor (red lines), red triangles: data use to fit the superexponential growth (blue lines). Right side: Time dependence of infected persons. Blue and red dots: empirical data, lines: results obtained from solving Eq. and (1.2) using the parameters from Table S1. The green points denote the daily doubling times, the full green lines denote the fit results including phase 2, dashed line is without phase 2.

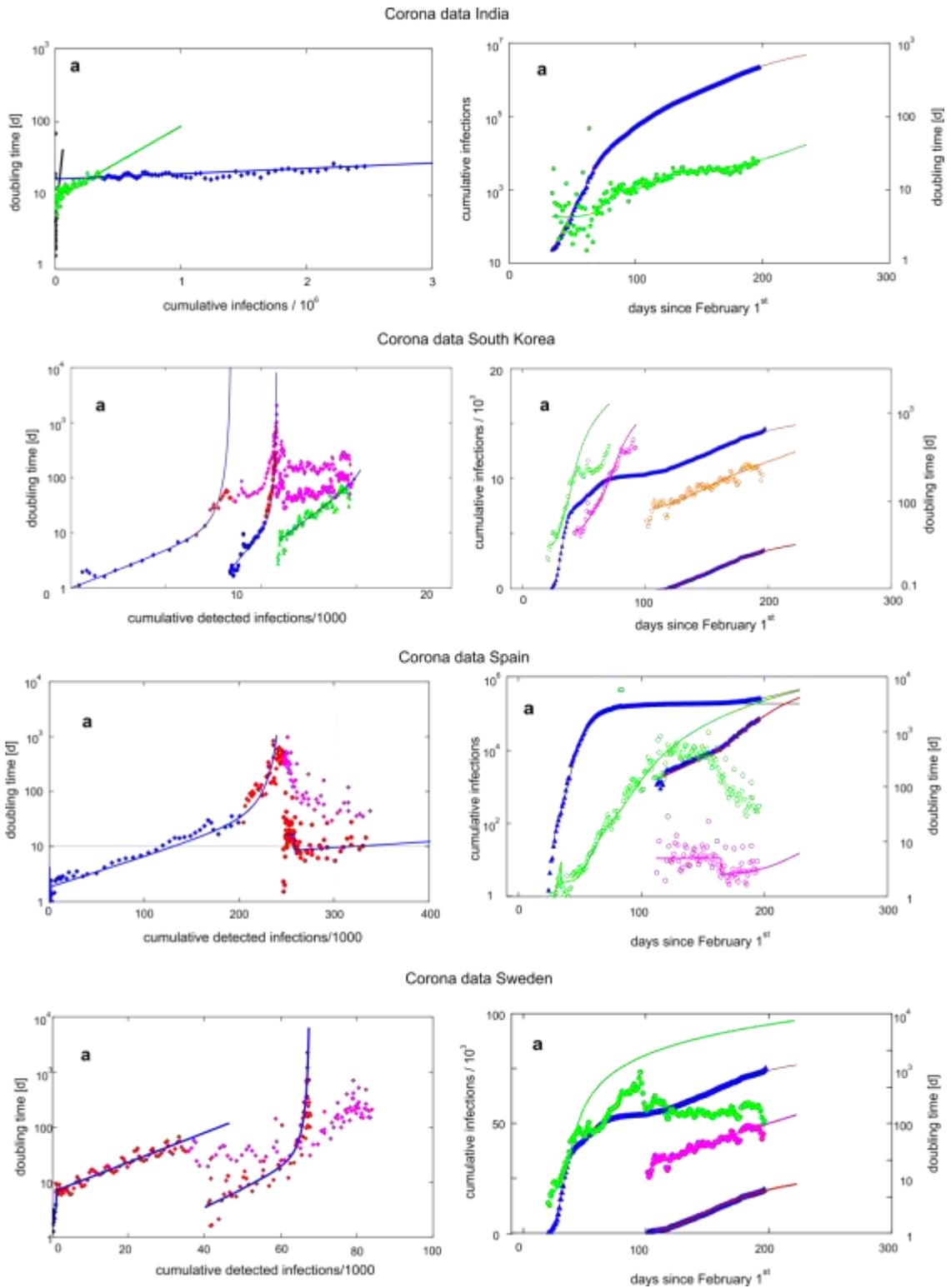

**Fig. S3:** same as Fig. S2

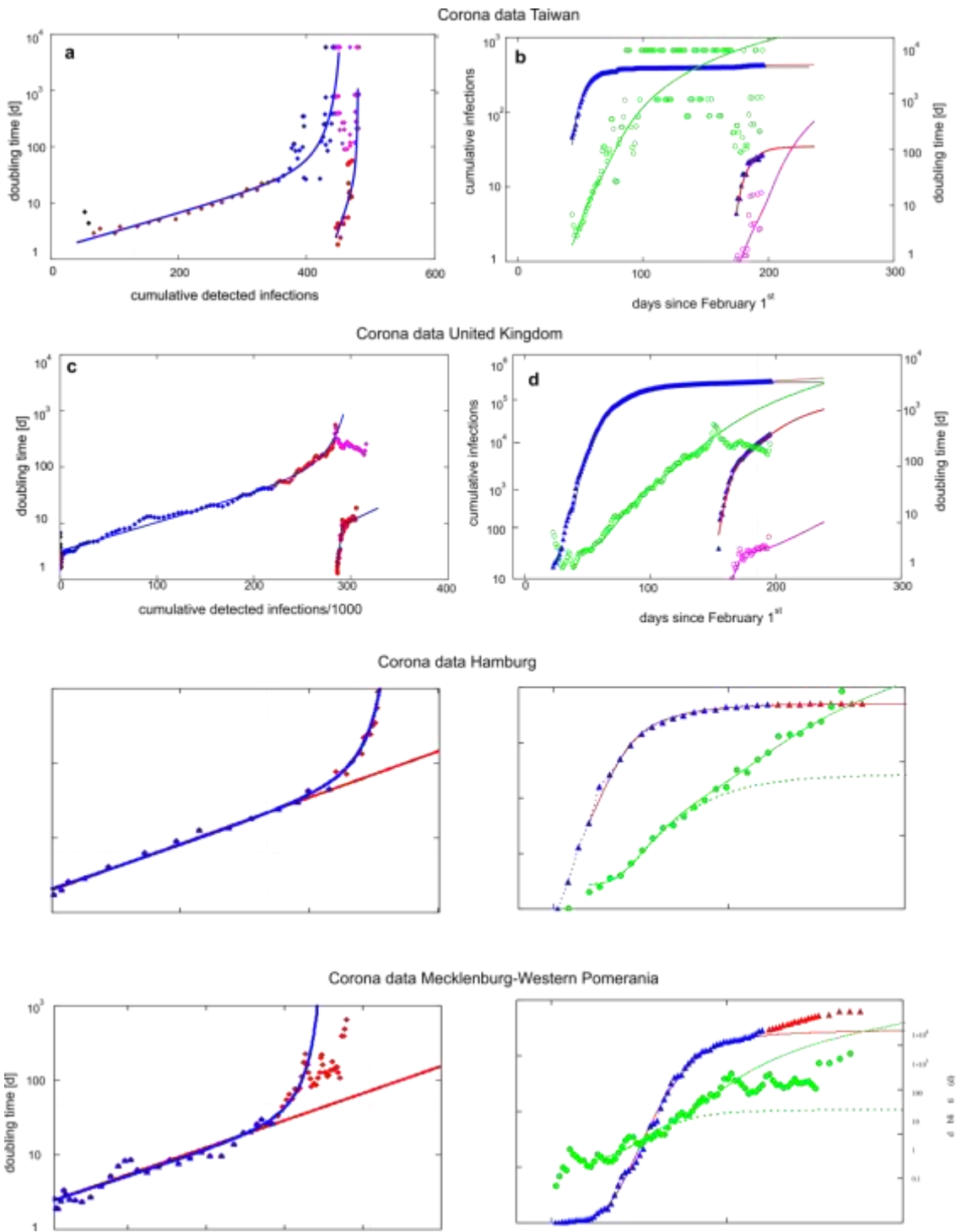

**Fig. S4:** Same as Fig. S2.

**Sections SI:** Analysis of Pandemic by SIR-based models

In order to compare our model to the hitherto used modelling we investigate four countries, namely Luxembourg, Switzerland, the Netherlands and Italy, which vary over a wide range in population numbers (see Table S1),

We used a simplified version of the SITAHRTE model of [SI1], which collects the different compartments of infected persons into only two, those infected but not confirmed, designated by I and those with a positive diagnosis (D=T+A+R+T), the other compartments being the same. This gives the name SIDHE for the model. In this way, the eight differential equations of the SITHARTE model are reduced to 5 and the number of parameters from 16 to 6 (see Fig. SI1). This model is similar to those used in [SI2,SI3] but we use a continuous time dependence of the parameters.

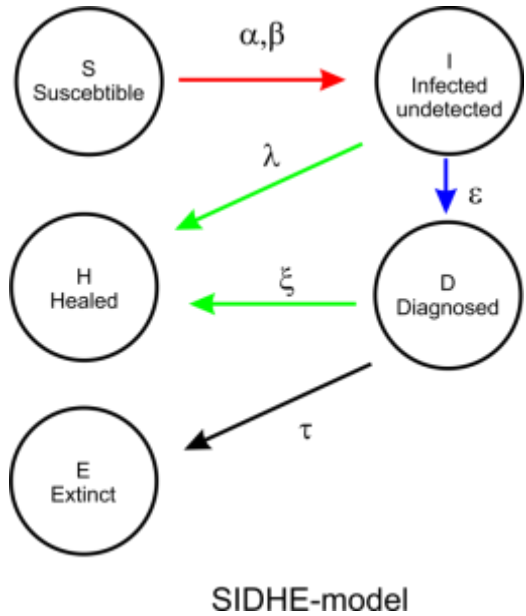

Fig.SI1: Graphical representation oft he SIDHE model used to describe the COVID-19 pandemic in a compartmental model

The model is described by the following five coupled differential equations

$$
\begin{aligned}
\dot{S} &= -S \cdot (\alpha I + \beta D) & (a) \\
\dot{I} &= S \cdot (\alpha I + \beta D) - (\varepsilon + \lambda)I & (b) \\
\dot{D} &= \varepsilon I - \xi D - \tau D & (c) \\
\dot{H} &= \lambda I + \xi D & (d) \\
\dot{E} &= \tau D & (e)
\end{aligned} \quad (1.1)
$$

We assume that the parameters $\alpha, \beta, \varepsilon$ can be time dependent, while the others remain constant. For the temporal dependences we assumed a maximum of three time ranges, the first comprising the first four to eigth days, where the pandemic spreads uncontrolled (with parameters $\alpha_0, \beta_0, \varepsilon_0$. The second time ranges comprises the first weeks of lockdown (up to day 20), the third range is up to day 60, then the pandemic spreads with the parameters from day 60 on.

To get a continuous time dependence we assumed for $f = \alpha, \beta, \varepsilon$ the following form

$$f(t) = (f_0 - f_2)\exp\left[-f_1\Theta(t-t_1)(t-t_1)\right] + f_2\exp\left[-f_3\Theta(t-t_3)(t-t_3)\right] \quad (1.2)$$

As shown below, this model allows describing the pandemic quite well with minimum set of parameters, for Italy similar to those given in [SI1], but also for the other examples.

We used a standard mean-square error minimizer base on the conjugated gradient method. We minimized both the cumulative cases $C(t)$ and the number of daily infections $\dot{C}(t)$ together. These quantities are given by the following expressions

$$C(t) = D(t) + E(t) + \int_0^t \xi(t')D(t')dt' \quad (a)$$
$$\dot{C}(t) = \varepsilon(t)I(t) \quad (b) \quad (1.3)$$

The results are given in Table SI1 for the four examples LU, CH, NL, and IT.

| Parameter | Luxembourg | Switzerland | Netherlands | Italy |
|---|---|---|---|---|
| Population (Mio) | 0.626 | 8.65 | 17.13 | 60.46 |
| α0 | 0.5013 | 0.507 | 0.528 | 0.4937 |
| α1 | 0.1639 | 0.0691 | 0.1341 | 0.0709 |
| α2 | 0.1515 | 0.0563 | 0.2372 | 0.0687 |
| α3 | 0.1028 | 0.0764 | 0.0103 | 0.0054 |
| β0 | 0.0263 | 0.01 | 0.0151 | 0.016 |
| β1 | 0.0981 | 0.0137 | 0.0996 | 0.0374 |
| β2 | 0.0058 | 0.001 | 0.0047 | 0.0012 |
| ε0 | 0.2606 | 0.142 | 0.1707 | 0.1137 |
| ε1 | 0.0583 | 0.0158 | 0.043 | 0.0124 |
| ε2 | 0.5053 | 0.4715 | 0.3454 | 0.1714 |
| λ0 | 0.0331 | 0.0089 | 0.0348 | 0.0073 |
| ξ0 | 0.0324 | 0.027 | 0.0155 | 0.008 |
| τ0 | 0.0005 | 0.0016 | 0.0049 | 0.0012 |
|  |  |  |  |  |
| T1 | 8 | 6 | 6 | 4 |
| T2 | 20 | 20 | 20 | 20 |
| T3 | 50 | 50 | 65 | 45 |
| D(0) | 20 | 18 | 24 | 22 |
| I(0) | 57 | 36 | 80 | 200 |
| $N_{inf}$ | 4174 | 32008 | 55848 | 252704 |

Table SI1: Parameters for the SIDHE model.

A closer look at Table SI1 reveals that no systematic behavior due to the still large number of parameters (in total 17, but much smaller than in the original SIDARTHE model, which for a three phase regime amounts to 53 [SI1]. Therefore, one tries to condensate their information into one quantity, the basic reproduction number $R_0$, which is given in the SIDHE-model as

$$R_0 = \frac{\alpha}{\varepsilon + \lambda} + \frac{\beta \cdot \varepsilon}{\xi \cdot (\varepsilon + \lambda)} \quad (1.4)$$

Its time dependence is shown in Fig. SI2, showing for all countries a similar behavior, but no obvious conclusion can be drawn. The same is true for the calculated endpoint of the pandemic, which growth with the number of inhabitants, but that is to be expected.

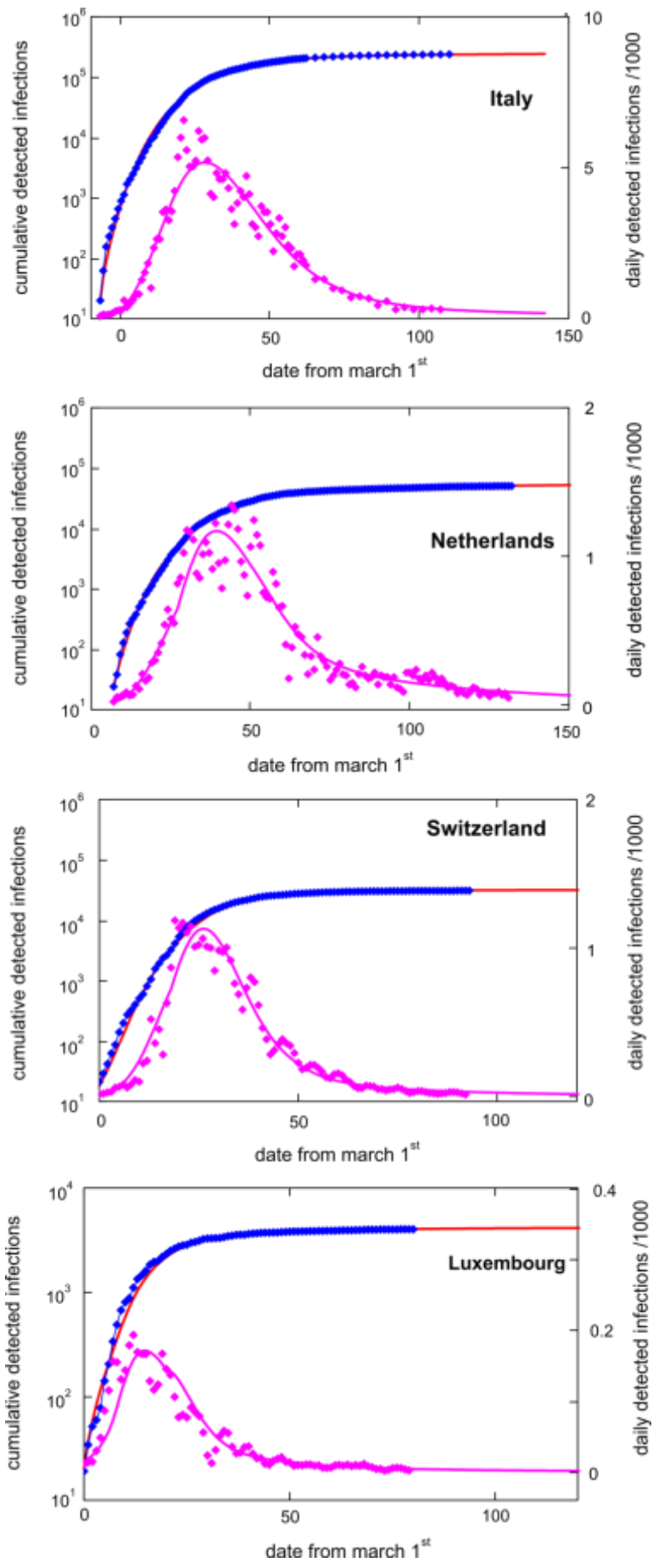

Fig.SI2: Solution of the SIDHE model for four different countries using the parameters as given in Table S1 (full red and magenta lines). Shown are both the empirical numbers for the cumulative detected infections (left abscissa, logarithmic scale) and the daily infection numbers (right abscissa, linear scale) taken from [3].

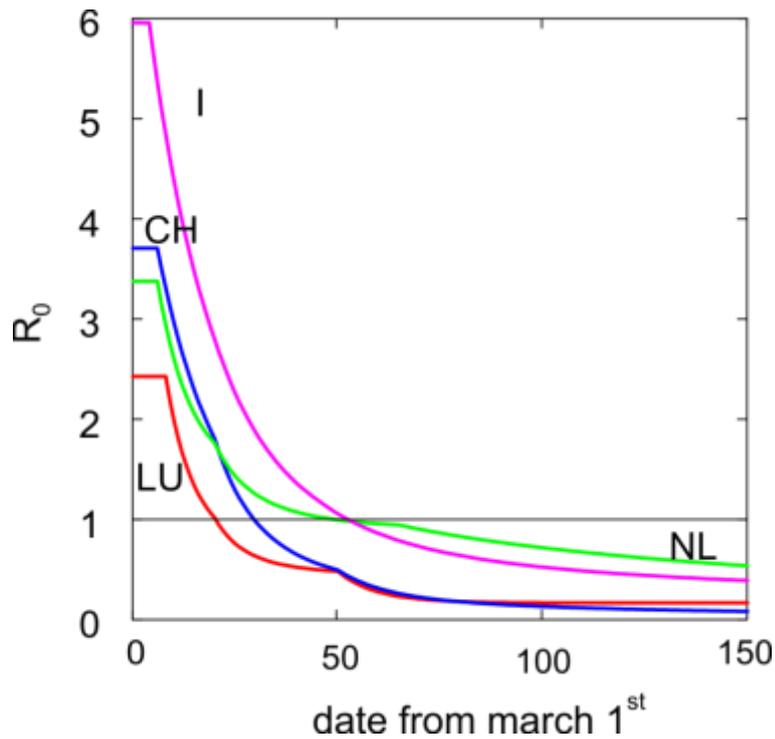

Fig.SI3: Basic reproduction number calculated from the SIDHE model for four different countries using the parameters as given in Table S1 .